\documentclass[12pt,preprint]{aastex}		

\shortauthors{de la Reza et al.}
\shorttitle{Complex organic and inorganic molecules in Li-rich K giants}

\begin{document}

\title{Complex organic and inorganic compounds in shells of Lithium-rich K giant stars}

\author{ Ramiro de la Reza\altaffilmark{1}, Natalia A. Drake\altaffilmark{1,2}, 
       Isa Oliveira\altaffilmark{1}, \& Sridharan Rengaswamy\altaffilmark{3,4} }

\altaffiltext{1}{Observat\'orio Nacional/MCTI, Rio de Janeiro, 20921-400, Brazil; email:
  delareza@on.br} 
\altaffiltext{2}{Saint Petersburg State University, Universitetski pr. 28, 198504 Saint Petersburg, Russia}
\altaffiltext{3}{European Southern Observatory, 3107, Alonso de C\'{o}rdova, Santiago, Chile}
\altaffiltext{4}{Current affiliation: C4C/115, Janak Puri, New Delhi, 110058, India}

\begin{abstract}

Hydrocarbon organic material, as found in the interstellar medium, exists in complex mixtures of aromatic and aliphatic forms. It is considered to be originated from carbon enriched giant stars during their final stages of evolution, when very strong mass loss occurs in a few thousand years on their way to become planetary nebulae. We show here that the same organic compounds appear to be formed in previous stages of the evolution of giant stars. More specifically, during the first ascending giant branch K-type stars. According to our model this happens only when these stars are being abruptly enriched with lithium together with the formation of a circumstellar shell with a strong mass loss during just a few thousand years. This sudden mass loss is, on an average, a thousand times larger than that of normal Li-poor K giant stars. This shell would later be detached, specially when the star stops its Li enrichment and a rapid photospheric Li depletion occurs. In order to gain extra carbon-based material to form the organic hydrocarbons, and also to explain the presence of complex inorganic compounds in these stars, we propose an interaction of these strong winds with remaining asteroidal/cometary disks that already existed around these stars since they were dwarf A-type stars. {\rm The mechanism of interaction presented here is successful to explain the presence of inorganic compounds, however it is unable to produce new carbon free atoms to form the organic hydrocarbon compounds.} Finally, we discuss some suggestions and speculations that can eventually help solving the long-standing puzzle of Li-rich giants.

\end{abstract}


\section{Introduction}
\label{sec:sec1}

Since the late seventies, organic compounds have been detected by means of unidentified 
infra-red emissions (UIE) in several cosmic environments, e.g., the interstellar medium (ISM), galaxies, post-asymptotic giant branch (AGB) stars, pre-planetary nebulae (PPN) and young protoplanetary disks \citep{sal11,gar11,cer09,bau09,ack10,li12}. 
Nevertheless, the mechanisms for the formation of these compounds are still obscure, and the nature of the carriers of UIE is still under debate \citep{kwo11, kwo13}.  
Are they free-flying gas molecules of polycyclic  aromatic hydrocarbons (PAH) or solid mixed aromatic/aliphatic organic nanoparticles (MAONs)?  
Hydrocarbons are very rarely detected in Oxygen-rich giant stars, except for some post-AGB binaries with circumstellar disks \citep{gie11} or some planetary nebulae \citep{guz14}.  
Among the first  ascending giant stars, aromatics were detected, for the first time, in the super Li-rich K giant HD~233517 \citep{jur06}.  
These authors invoked a hypothetical recently formed disk to explain the presence of aromatics. 
In this paper, we supplement our model of connecting Li enhancement and infrared excesses \citep{del96, del97, del12}  with the Spitzer mid-infrared spectra of seven Li-rich K giants. 
{\rm In addition to the hydrocarbon organic compounds, we report here the presence 
of peculiar inorganic compounds in the spectra of first ascending K-giant stars. 
In fact, these spectra are completely different from those of normal K giants \citep{slo15}.}
We propose that Oxygen-rich K giants, that are also momentarily Li-rich, can eventually obtain new carbon by partially perturbing disks around them, and thus form hydrocarbons. 
These disks are the debris disks that have existed since the stars  were $\sim$A5 dwarfs \citep{del14}. 
As they suddenly become Li-rich, owing probably to the processes that are described in Section \ref{sec:sec4}, there is a mass loss ejection of dust and gas that is, on an average, a thousand times larger than that occurring in a normal K star in the first ascending giant branch. This mass loss leads to the formation of a circumstellar shell. 
In \citet{del12} it is proposed that this shell emerges from the stellar surface and remains attached to the star during the Li-enrichment phase but gets detached and 
ejected when the Li-enrichment ceases. 
 The photosphere of the star with a detached shell depletes rapidly, attaining the normal low Li abundance of K-giant stars. This fast Li-rich episode occurs at the rate of ~1\% of Li-rich K giants in the Galaxy \citep{del12, kum11} and thus appears to be a universal phenomenon. This means that all K-giant stars could pass by this short Li-rich episode, since the same rate has been found in other galaxies \citep{kir12}.

{\rm The proposed shell-disk interaction model is used to explain the several features of 
inorganic origin of K giants shown in this work. }          
This is the first time, to our knowledge, that a collection of emission line features superposed on a strong continuum emission (see Section \ref{sec:sec2}) is detected in first ascending giant stars. 
These spectra were discovered in the {\it Spitzer Space Telescope} archive, and our main motivation in this work is to discuss possible reasons for their presence.

\section{The Observed Mid-IR Spectra}
\label{sec:sec2}

\subsection{Observations}

\begin{figure}                           
\includegraphics[angle=0,scale=.7]{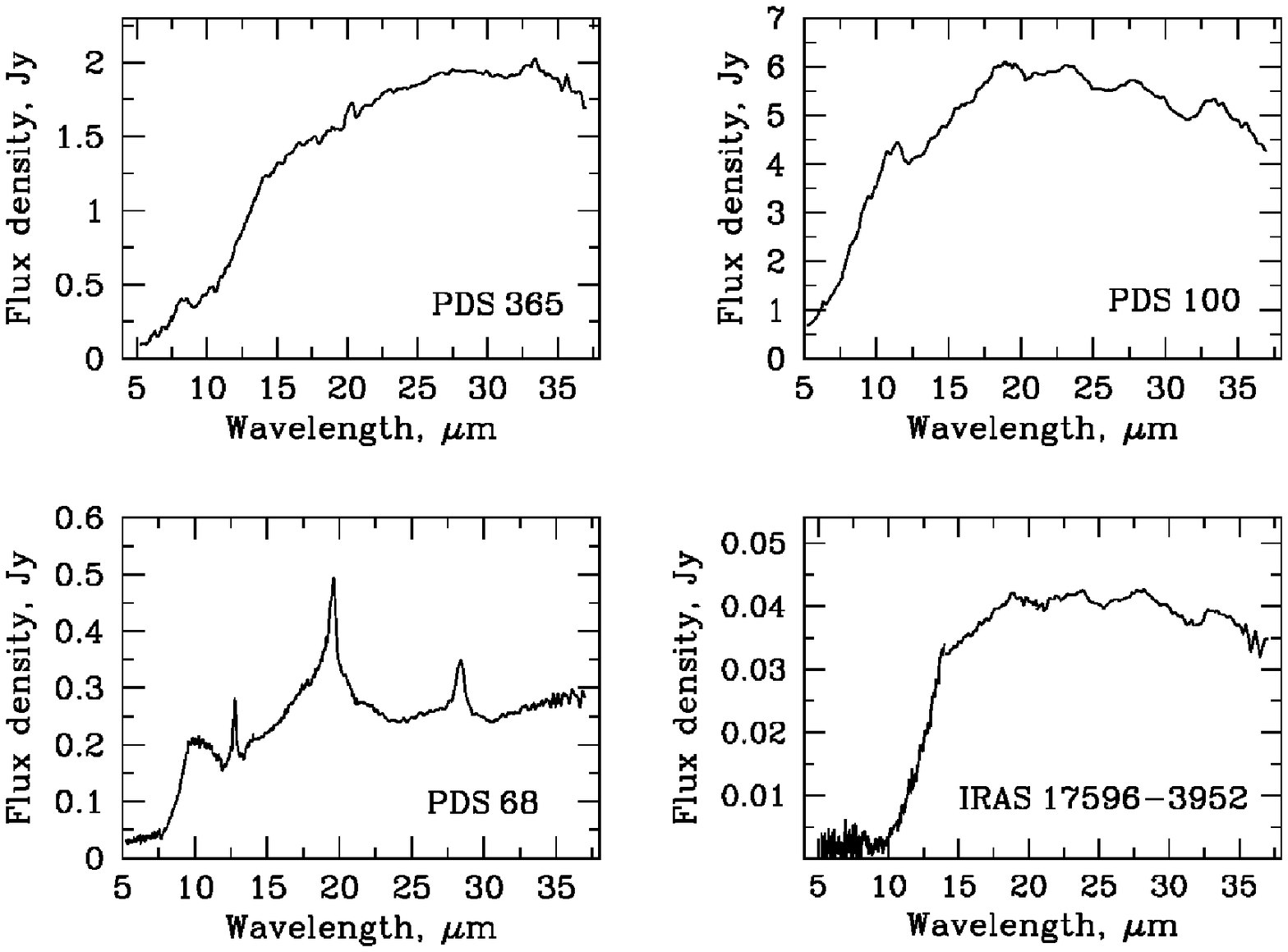}
\caption{Mid-IR spectra between 5 and 38~$\micron$ of RGB Li-rich K giants PDS\,365, PDS\,100, 
PDS\,68, and IRAS\,17596-3952 presenting the UIE features superposed to a strong adjacent 
continuum emission.\label{fig1}}
\end{figure}

\begin{figure}                           
\epsscale{0.8}
\includegraphics[angle=0,scale=.5]{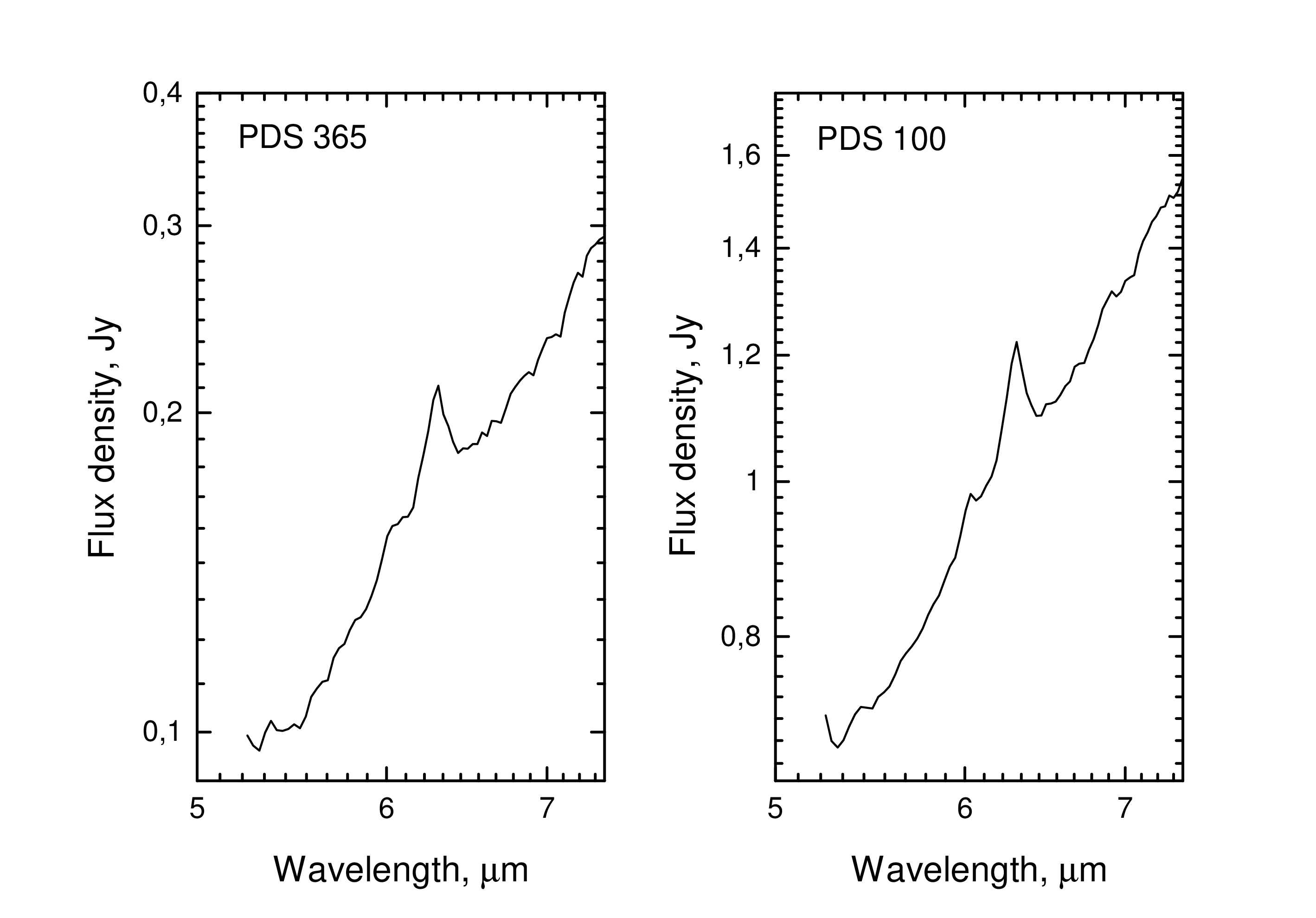}
\caption{A log-log form is presented here for stars PDS\,365 and PDS\,100 to highlight the 
emission feature at 6.26~$\micron$.\label{fig2}}
\end{figure}

This work is based on the mid-IR spectra of seven K-giant stars. 
Five of them are first ascending giant stars (RGB) which are: HD\,233517, PDS\,365 (IRAS\,13313-5838), PDS\,100 
(IRAS\,19285+0517), PDS\,68 (IRAS\,13539-4153), and IRAS\,17596-3952. 
The two remaining stars, IRAS\,12327-6523 and IRAS\,17582-2619, can be classified as early-AGB stars. 
Two other K-giant stars HD\,30834 and HD\,146850 are also considered, for comparison purposes. 
All spectra shown here were obtained from archives of the Cornell Atlas of Spitzer/IRS sources (CASSIS) \citep{lebou11}. 
All the spectra were observed in the wavelength interval of 5 to 38~$\micron$, with the exception of IRAS\,17582-2619 that shows data from 5 to 13~$\micron$ only. 
The spectra of four RGB stars, PDS\,365, PDS\,100, PDS\,68, and IRAS\,17596-3952, all of which 
have the property of being Li-rich or super Li-rich giants, are presented in Figure \ref{fig1}. 
A similar spectrum of HD\,233517 can be seen in \citet{jur06}. 
In Figure \ref{fig2}, an even more detailed parts of the spectra of PDS\,365 and PDS\,100 are presented, showing the spectral emission feature at 6.26~$\micron$. 
The spectra of the two early-AGB stars are presented in Figure \ref{fig3}, and the structureless spectra of the two comparison Li K-giant stars HD\,30834 and HD\,146850 are shown in Figure \ref{fig4}. 
All spectra have been observed in 2006 except that of IRAS\,17582-2619. 
The latter has been observed in 2007. CASSIS considers the sources as point-like, except for PDS\,365 that is taken as an extended source of $\sim$2.0$\arcsec$ and IRAS\,17596-3952 
($\sim$3.2$\arcsec$).  
Appropriate best extractions have been applied individually by CASSIS. 
We performed order and flux matching (for different aperture sizes) to the extracted spectra. 


\subsection{Stellar properties and compositions}   

Some of the main stellar properties of the seven Li-rich K giant stars are presented in 
Table~1. 
Out of the five RGB stars, three are located at the luminosity bump and the other two 
(PDS\,68 and HD\,233517), the two super Li-rich stars, are located very near to this 
luminosity bump region. 
This is the region where the giant star locally slows down its evolution and {\rm it corresponds 
to} stellar luminosities and effective temperatures of $\log (L/L_\odot) = 1.45 - 1.90$ and 
$T_{\rm eff} = 4450 - 4600$~K \citep{kum09}. 
The remaining two stars are early-AGB stars, having the lower limit of Li abundance of 
Li-rich giant stars. 
The Li abundance $\log \epsilon$(Li) is measured with respect to the number of hydrogen atoms with $\log \epsilon$(H) = 12.\footnote{$\log\epsilon({\rm Li})=\log(N_{\rm Li}/N_{\rm H}) +12$.   
The Li abundance in the interstellar medium is $\log \epsilon$(Li) = 3.2.} 
The standard theory of stellar evolution predicts a maximum of $\log \epsilon$(Li)=1.4 for the RGB phase of stars as the Li is expected to be depleted {\rm \citep{ibe67, ibe91}}.

{\rm In Table~1, together with some important stellar parameters, we have included the most important spectral UIE features with their respective and most probable aromatic, aliphatic and inorganic identifications. As an example, we show in Figure \ref{fig2} the aromatic features at 6.26~$\micron$ of two stars from this work. }
In general, aromatics features appear as narrow structures and aliphatic plateaus are broad in mid-IR spectra \citep{kwo11,kwo13}. 
Also, a new condition is observed here, with aromatic and aliphatic compounds coexisting and developing in the absence of {\rm a strong UV radiation}. 
We find that UIE features appear in the mid-IR spectra, as it is the case of RGB giants, only in those stars that have a strong continuum emission, in agreement with the general cases mentioned by \citet{kwo13}. 
This continuum emission is clearly seen in all spectra of the Li-rich RGB stars, from 
5 - 10~$\micron$ up to 38~$\micron$ as shown in Figure \ref{fig1}. 

{\rm As mentioned in Section \ref{sec:sec1}, two problems are faced here. 
One is that of the origin of the organic compounds, that are considered to be formed a priori in the shells. 
The other is the origin of the inorganic compounds which, at least for the five RGB stars, could be the result of the wind-disk interaction that will be discussed in Section \ref{sec:sec3}. }

To our knowledge, no organic compounds have been found in debris disks even in young objects such as HR\,4796 A \citep{koh08}, but only in gas rich protoplanetary disks. 
We then consider that the organic compounds are probably formed in the shell winds. The UIE phenomena is a complex one \citep{kwo13} and we are still far from understanding the detailed formation of these UIE features corresponding to the organic compounds presented in Table 1. Identifying the carriers of these UIE and selecting one of the two scenarios mentioned in Section~1 is the first step towards making progress in this field. 
Are these UIE carriers 2D free-flying PAH molecules? 
Or are they 3D solid structures like amorphous nano particles (MAON) containing a much larger number of carbon atoms as those of PAHs? 
This subject has been largely discussed in \citet{kwo13}. 
In this work, we find two important observed properties that are in favor of the MAON scenario in agreement with the conclusions of \citet{kwo11,kwo13}. 
One of them is the strong correlation found here between the presence of UIE in the RGB giant's spectra and the presence of a strong underlying continuum in emission. 
This is because this continuum can only be due to large (micron-size) solid state particles. 
The second property is that we detected 8 and 12~$\micron$ plateau features (see Table~1) that, according to \citet{kwo13}, are of aliphatic origin and would have no place in the PAH scenario.

All these evidences lead us to propose that the mechanism for formation of the organic compounds in the emerging shells of Li-rich K giant stars has some similarities to those found in carbon-rich PPN. 
These similarities are: (i) the presence of a negligible UV continuum radiation in both cases; and (ii) that this phenomenon requires similar formation time scales, of the order of some few thousand years. 
The differences are that in the case of the PPN, stellar mass losses are two orders of magnitude larger than in the case of these Li-rich K giant stars. 
Also, carbon particles are intrinsically part of the winds in case of carbon-rich PPN whereas in our scenario of Li K giant stars, carbon is supposed to, in principle, come from the wind shell interaction with disks.


\section{The Shell-Disk Interaction Model}  
\label{sec:sec3}

{\rm In the nineties Greg\'orio-Hetem et al. (1993) realised that a large majority 
of Li K-giant stars were IR sources measured by IRAS. 
In a diagram based on fluxes at 12, 25, and 60~$\micron$, they found that these sources 
were distributed in mainly two regions. 
One of them, formed by faint visual stars characterised by excesses at 25~$\mu$m  and 
the other one, formed by bright stars with excesses at 60~$\mu$m. 
There is also a third region constituted by normal Li-poor K-giant stars without IR excesses. 
In  \citet{del96, del97} we constructed a simple model connecting 
those three regions. 
This model considered that departing from the non-IR excess region, each star being 
recently and abruptly enriched with $^7$Li, is related to the ejection of a shell of gas 
and dust from the star to the ISM. 
In this way, the shell near the star radiates at 25~$\mu$m whereas far away, radiates at 60~$\mu$m. 
When the shell is completely ejected, the star returns to the origin closing the loop of these changes. 
In \citet{del12} we have introduced a more realistic description of the emergence of the shells. 
In Figure~5 we present this model and its figure caption gives more details on the model. 
With the selected values in the figure, a complete loop is realised in $\sim$80\,000 yr. 
This model can be verified either by directly imaging these shells or by analyzing their spectra 
at corresponding wavelengths.
This is what we are trying to do in the present work by means of spectra centered at $\sim$25~$\mu$m.}

{\rm Let's} invoke a wind shell-disk interaction scenario to explain the presence of 
complex inorganic compounds observed in the shells, and to examine how to obtain extra 
free carbon based material, eventually necessary to produce the organic compounds. 
Because we are in face of short timescales, but very strong winds, with mass losses about one thousand times larger than those of normal RGB stars, we only consider here the stellar wind drag acting on a debris disk, supposedly the residual disk from when the K-giant star was an A-type star in the main sequence.

The emerging shells, supposedly spherical and attached to the star at this stage, 
{\rm could result from a sudden internal transport of matter connected with an also sudden $^7$Li 
enrichment of the stellar surface (see Section 4).
Following our model, this is represented by the stars labelled in red in Figure \ref{fig5}.}
These circumstellar shells can reach distances of the order of 800 AU, with a minimum expansion velocity of 2~km\,s$^{-1}$ during 2000~yrs {\rm \citep{del12}}. 
Then, any body $b$ of a disk rotating around the central star will suffer, in principle, a drag during the time period which is the lifetime of the shell. 
It is only during this period that the radiation emitted by the shell is concentrated in the mid-IR region, the explored spectral region of this work.

If we consider a star with a mass $M_S$  and a shell formed by a mass loss $\dot{M}_{S}$, the gas/dust wind density will be  $\varrho_W$ = $\dot{M}_{S}$/4$\pi\,D^2V_W$  where $V_W$ is the wind velocity and $D$ the distance of body $b$ in the disk to the central star. 
Let us consider now the encountered rate mass $M_e$ \citep{jur08} defined as the rate of shell mass encountered by an spherical body $b$ in the disk with radius $R_b$  and density $\varrho_b$. 
This encountered rate is defined by $\dot{M}_e = (\pi\,R_b^2)\varrho_b\,V_b$, where $V_b = (GM_S/D)^{1/2}$.  
By replacing $\varrho_W$ and $V_b$ and integrating over the wind lifetime $\Delta t$, we obtain:

\begin{equation}
M_e \approx R_b^2(\dot{M}_S/4) V_W^{-1} D^{-5/2} G^{1/2}(M_S)^{1/2} \Delta t 
\end{equation}
                                                                                                           
If $M_e$ of the shell is equal to the mass of the body, this can suffer the drag, lose momentum and will spiral towards the star. In this case we have $M_e = 4 \pi\,\varrho_b R_b^3/3$, which enables us to estimate the critical radius of the body suffering the drag as:

\begin{equation}
R_b^C \approx (3\,\dot{M}_S/16\pi)V_W^{-1} D^{-5/2}\varrho_b^{-1} G^{1/2}(M_S^{1/2})\Delta t
\end{equation}
                                                                                                              
Any body with a radius larger than  $R_b^C$ will not suffer the drag and will survive at the corresponding distance $D$. 
In this scenario, it is expected that a debris disk of a dwarf A-type star entering into the post-main-sequence phase could suffer important transformations in its structure. 
In fact, detecting debris disks in the RGB phase is a difficult task \citep{bon10}. 
This is probably due to their less dusty material. 
More success can be obtained when observing sub-giants stars, specially those hosting planets.

Typical main sequence debris disks appropriate for our case are those spatially resolved corresponding to A3-A6 stars, as those belonging to the star $\beta$ Pictoris \citep{van10} and HR\,8799 \citep{su09,mat14}. These disks have an internal radius smaller than 10~AU and an external radius near 2000~AU. 
We can now estimate the value of the critical radius $R_b^C$ for our case by considering typical values as those presented in Figure \ref{fig5}. 
We have $\dot{M}_S = 10^{-7}M_\odot$ yr$^{-1}$,$V_W = $ 2 km s$^{-1}$, $M_S\sim 1.0-1.5M_\odot$. As discussed above, we expect to find a debris disk for the RGB formed mainly by asteroids with few dusty material. For $\varrho_b$, we then chose a value of 2.1~g\,cm$^{-3}$ typical of a large asteroid \citep{mic00, jur08}. 
We select a mean value of $\Delta t $= 1500~yr and for $D$ = 10, 50, 100, 500~AU we obtain approximate values for $R_b^C$ of 2.2, 0.04, 7$\times$10$^{-3}$ and 1.3$\times$10$^{-4}$~cm, respectively, representing small particles. 
Radiation pressure in the 
giant phase will remove objects with diameters smaller than 0.01 cm, as can be seen in figure~6 of \citet{bon10}. 
In our case, only the particles with radii of 2.2 to 0.04~cm will not be removed by radiation pressure. 
We have also to take into account that the dragging mechanism considered above is only efficient when it acts in the inner regions of the disk. 
For example, at $D$ = 10~AU the orbital time of an asteroid is 25 times smaller than the wind lifetime of  $\sim$1500~yr used here. In other words, for short $D$ distances, the adiabaticity of the mass loss is approximately maintained \citep{bon10}. 
For larger $D$ distances this adiabaticity breaks-down and dragging becomes inefficient. 
 
In order to show how our proposed dragging mechanism can explain the presence of the observed inorganic compounds, lets consider the case of the super Li -rich RGB star PDS 68 in Table 1 and also Figure \ref{fig1}. 
Three strong emissions features are present at 12.7, 19.5 and 28.3~$\micron$. The distribution of these three emission lines coincides those of crystalline Enstatite with dimensions of 1.5 $\micron$ (see figure 9 in \citet{olo09}). 
In fact, these particles are found in the inner parts of disks \citep{bou08,juh10} where our dragging mechanism is efficient. These micron size particles have been maintained and not removed in the expanding shell, during these very short times, of the order of 1500 yr, during which these stars are detected. 
It is {\rm then} expected that the observed continuum emission on the RGB stars is due to micron-size solid particles that are also maintained in these short lifetime shells. 
What is not clear to us at present is how 
to obtain fresh new carbon from this shell disk interaction in order to help the formation of organic hydrogenated carbons. 
It must be first noted that if the disk of $\beta$ Pictoris contains a gas component rich in CO \citep{cat14, den14}, it is due to the very small age of 11 Myr of this star \citep{ort04}, even if the origin of this gas is not completely clear. It is possible that at the stage of our RGB stars, CO gas can be liberated from cometary material by means, for example, of electrons and H$_2$O collisions \citep{mat15} . 
Nevertheless, once the CO gas is liberated, we find no conditions in our scenario to produce the photo-dissociation of this molecule, as apparently is the case  of the oxygen-rich nebulae, in order to form aromatic compounds \citep{guz14}. In our case, the star radiation is too low.

\begin{figure}                              
\includegraphics[angle=0,scale=0.7]{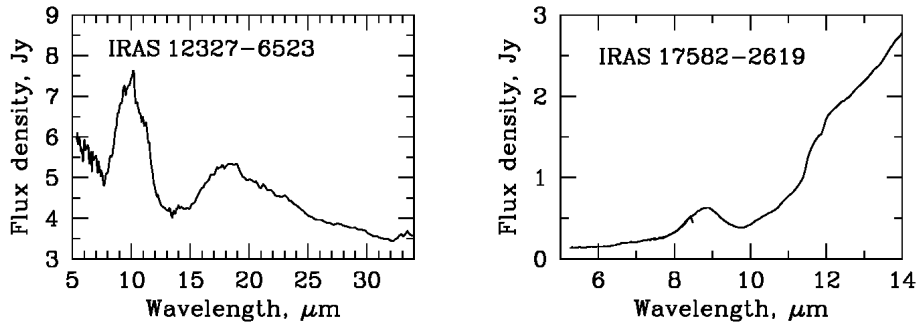}
\caption{The mid-IR spectra of the only two early-AGB stars where emission continuum 
is much less evident.\label{fig3}}
\end{figure}

\begin{figure}                            
\includegraphics[angle=0,scale=.7]{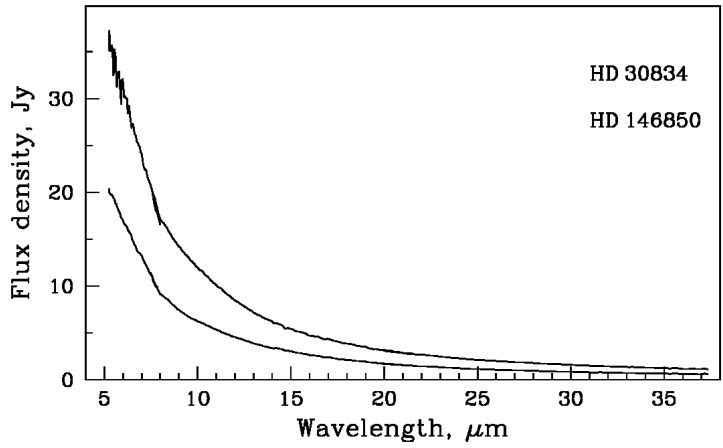}
\caption{These are the mid-IR spectra of the two stars labelled in blue in Figure \ref{fig5}. 
The two Li K-giants HD\,30834  and HD\,146850 do not exhibit continuum emission and the associated UIE features.
The different shape of these two spectra in respect to those presented in previous figures 
is explained in Section~3.\label{fig4}}
\end{figure}

As concerning the two objects here classified as early-AGB stars in Table 1, IRAS 17582-2619 \citep{gar97} and IRAS\,12327-6523, they have mass losses of $10^{-7}M_\odot$ yr$^{-1}$ and $5\times10^{-8}M_\odot$ yr$^{-1}$ respectively, following Figure \ref{fig5}. Star IRAS\,12327-6523 appears not to present a strong continuum emission typical of RGB stars. Its spectrum contains two strong emissions at 10 and 18.5 $\micron$ (Figure \ref{fig3}, left panel) of possible amorphous inorganic material that could be originated in its own stellar wind. 
Because of its low mass loss, it is expected that crystallization is inefficient due to the lower dust formation temperatures expected for these low-mass rates \citep{suh02}. 
The other star, IRAS\,17582-2619, even with a short observed spectral interval, appears to show an increasing emission continuum (Figure \ref{fig3}, right panel). 
We have tentatively identified the emission at 8.7~$\micron$ as an aliphatic plateau.

{\rm All the shell spectra discussed up to now, represent emerging shells attached to the stars which are represented by red points in Figure 5. Nevertheless, the  spectra of the two Li K-giant stars HD 30834 and HD 146850 shown in Figure 4 are completely different. They are characterised by the absence of a continuum in emission and also by the total absence of  UIE features. The shapes of these structureless spectra are similar to those of ordinary K giants \citep{slo15}. 
The positions of these two objects in Figure 5 (blue points) indicate for our model, that their shells are already detached, leaving free the zone that would emit at $\sim$25~$\mu$m and emitting only at 60~$\mu$m far away from the star.}

Finally, it is important to mention that the shell-disk interaction can introduce a 
non-spherical geometry in our scenario and show important optical polarisation signals. 
In fact, two of our sources, PDS\,100 and PDS\,68, exhibit a large degree of polarisation \citep{per06}.

\section{Some Considerations on the Li Enrichment Processes}
\label{sec:sec4}

More than thirty years have elapsed since the discovery of the first Li-rich K-giant star and no self-consistent models have appeared explaining this Li enrichment phenomenon excluded by the standard theory of stellar evolution. Stellar internal or external scenarios have appeared trying to solve this ``Li rich puzzle". 
Concerning the external approach based in general on the engulfing of a planet \citep{sie99}, strong observational evidences have worked against this scenario, as the absence of a simultaneous $^9$Be and $^6$Li enrichment together with $^7$Li.  
Also, due to the gain of planet momentum, these stars would be predominantly rapid rotators which is not the case.  
As far as the internal models are concerned, the most appropriate for the cyclical model discussed here, the situation is the following: all scenarios are based on the Cameron-Fowler $^7$Be mechanism \citep{cam71} to create new $^7$Li. 
However, all low mass stellar models, as is the case here, require a rapid outward 
transport of the $^7$Be in order to avoid its destruction before it could decay into $^7$Li in the cool external convective envelope (CE) and enrich the stellar surface with fresh $^{7}$Li.
Theoretical difficulties arise by the fact that, in the absence of a rapid mixing mechanism, the models respond well, explaining the enrichment of the other elements such as $^{13}$C and also the normal $^7$Li depletion. However, different 1-D evolution codes used for the $^7$Li depletion seem to produce different Li abundances \citep{lat14}.  The two scenarios, a rapid and a slow mixing, are then somewhat opposite. To conciliate them, the rapid process must be concentrated in short evolutionary episodes in an otherwise normal slow and long evolutionary process. This is what we propose in this work.

\citet{egg08} detected the appearance of an instability relating the top of the H-burning zone and the bottom of the CE by means of 3D stellar models of low mass giants. This non-convective mixing is produced by the reaction $^3$He ($^3$He,2p) $^4$He on top of the H-burning zone inducing a molecular weight inversion. This happens precisely at the LB where we found all the RGB Li rich K-giants reported in this paper. This mixing process is supposed to destroy a large part of the $^3$He. However, if under certain short timescales, part of $^3$He is used to create $^7$Be by means of $^3$He($^4$He,$\gamma$)$^7$Be and rapidly transported to the CE before being destroyed, a net $^7$Li enrichment can be obtained at the stellar surface by the decay $^7$Be (e, $\nu$)$^7$Li. 
Nevertheless, the velocities of 1 ms$^{-1}$ proposed by \citet{egg08} are probably not large enough to produce the required excess of $^7$Li. At this point, we can speculate and explore a new mechanism for this rapid transport. This is achieved by 
relating this scenario to the internal angular momentum (AM) loss necessary to reduce the stellar core rotation of these giant stars. In fact, recent asteroseismology measurements of the {\it CoRoT} and {\it Kepler} satellites have shown that at the sub-giant stages, the stellar cores appear to suffer a spin-up \citep{deh14}. This spin-up is transformed into an important core spin-down at a later stage when the stars advance in the RGB \citep{mos12}. These low core rotations appear at the upper RGB and in the He-burning clump giants \citep{can14}. These two regions are where Li-rich K-giants exist.  We speculate that the internal AM transport mechanism --whatever it is-- can act in a continuous way but has peaks at the LB in order to produce the Li-enhancement, while the gain of AM by the envelope could eventually produce the observed shells presented in this work.

As far as the more massive (2$M_\odot$) clump Li-rich giants are concerned (see for instance, 
\citet{car14}) the phenomenology can be different because these stars, having already experienced the He flash, can eventually generate Hydrogen Flashes \citep{sch01} that could be involved in the $^7$Li creation and contribute to the reduction of the core rotation velocities.

\section{Results and Conclusions}
\label{sec:sec5}

\begin{figure}
\includegraphics[angle=0,scale=.5]{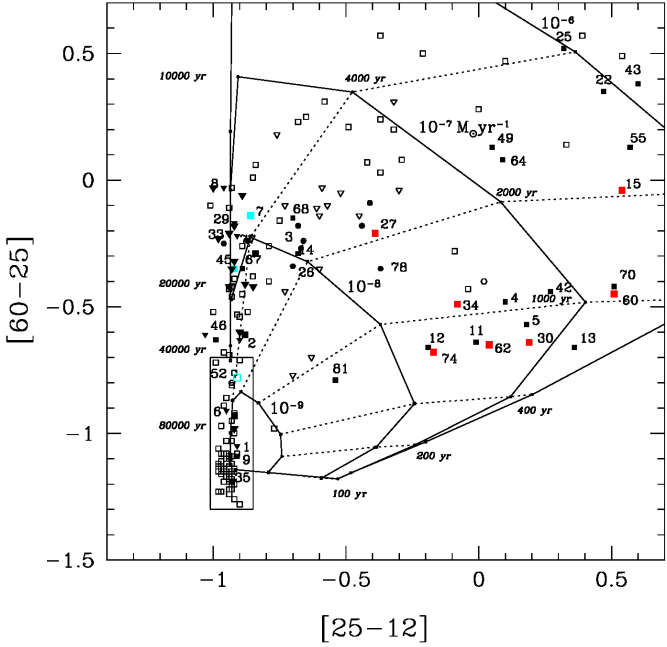}
\caption{{\scriptsize Distribution of far-IR excesses of K giant stars: 
Among the very large number of red giant stars observed  by the IR Astronomical Satellite (IRAS), more than hundred K-giant stars present IR excesses in fluxes at 12, 25 and 60~$\micron$. 
In this figure, the Li-rich (filled square symbols) and Li-poor (open symbols) K giants and those with upper limit values (inverse triangles) are shown.  
Observed Li stars presenting UIE features considered to be in the new attached emerging shells are labelled in red with the same numbers as in the original model figure in \citet{del97}. 
Labelled in blue are the stars with considered detached shell where no UIE appear. 
Four shell evolution curves are presented with their respective time steps calculated for typical stellar temperature $T_{\rm eff}$=4000~K and stellar radius of 20\,$R_\odot$. 
They correspond to different stellar mass loss values between 10$^{-9}$ $M_{\odot}$~yr$^{-1}$ up to 10$^{-6} M_{\odot}$~yr$^{-1}$ and $V_W = 2$~km\,s$^{-1}$. 
All kind of stellar photospheric rotational velocities are represented here even with a smaller fraction of rapid rotators \citep{dra02}.  
Here the axis are defined  by [$\lambda_2-\lambda_1] = \log (\lambda_1S_{\lambda 2})-\log ( \lambda_2S_{\lambda 1}$) where $S_{\lambda}$  is the density flux.  
In general, at the right part of the diagram, where the shells labelled in red are located, 
belong to faint and distant K giants which have been detected as a sub-product in the PDS survey dedicated to search T Tauri type stars \citep{gre92,tor95}. 
On the contrary, objects at the left part, are bright and nearby K giants. 
Due to this different space volume in the Galaxy, we were able to detect in the right part, generally more rare K giants presenting rapid processes as is the case here of the recent emergence of a shell characterized by an excess at 25~$\micron$ in times of 1000 to 2000~yr following the model. 
Also, because of this, the nearby shells at the left, show longer phenomena up to 80\,000~yr. 
Here these shells appear detached with excesses only at 60~$\micron$. 
Note that at the shell emergence region at the right of the diagram, almost all objects are Li-rich. 
This is not the case at the left, where exist several Li-poor K giants rapidly Li depleted following the model, with a vanishing shell.}\label{fig5}}
\end{figure}

We detected several emission lines related to hydrocarbon organic and other inorganic materials in the {\it Spitzer} Mid-IR spectra of seven K-giant stars. Five of them are first ascending giants and two are apparently early-AGB stars. 
These emission features appear, specially in the case of the RGB stars, superposed to a strong underlying continuum emission between 5 and 38 $\micron$. 
To our knowledge, this is the first time, at least for first ascending giant stars, that this collection of continuum and lines in emission is found. 
Concerning the five RGB stars, all of them are Li-rich or super Li-rich, even reaching very high Li abundance values ($\log \epsilon$(Li) = 4.3). 
Also, all of them are located at or very near the luminosity bump. We suggest that the continuum emission in the RGB stars is due to an emerging circumstellar shell attached to the star that is associated with the rapid Li surface enrichment. 
This emergence process, as indicated in our model represented in Figure~\ref{fig5} by stars with red labels, occurs in very short episodes (of the order of 1000-2000 yrs). 
A general, internal stellar evolutionary process that can eventually be at the origin of this rapid Li enrichment together with a circumstellar shell formation, is discussed in Section \ref{sec:sec4}.

The hydrocarbon organic compounds, consisting of aromatic and aliphatic components, are supposed to be formed in the winds of the expanding shell. The details of this formation are however unknown, but we consider that they could be formed in a similar way as in carbon-rich pre-planetary nebulae. This is because the time scales are similar in both cases and also both evolved without the presence of a strong UV radiation field. Further, in contrast to the carbon rich PPN scenario, we are here in a case of dual chemistry in which hydrocarbons are formed in an oxygen-rich wind. 

We postulate a priori that a source of fresh carbon, to supply the formation of organics in the shell, could be the result of shell wind dragging on a debris disk surviving from the main sequence stage when these giants were A-type dwarfs. 
We are unable however, to find a physical process that can furnish free new carbon in our shell-disk scenario. However, because these organic features are observed, this leaves an open possibility that the used carbon could be that of the original star itself. This problem can possibly be solved in the future, when the carriers of organic UIE features will most be definitively discovered. These are 2D free-flying PAH molecules or 3D solid mixed aromatic/aliphatic nano particles (MAON). In this work we found properties supporting the MAON scenario by observing pure aliphatic plateaus and, mainly, by finding an strong correlation between UIE features and a strong underlying emission continuum in the five RGB stars. 
This continuum emission would be due to the emission of micron-sized solid state particles forming the circumstellar shells according to the MAON scenario.

Nevertheless, the action of this shell-disk interaction is successful in explaining the presence of the observed inorganic compounds in the shell. This is due to the effect of the dragging action on expected disks in the RGB phase. 
Due to the short (1000 - 2000 yr) action of a strong wind, correlated to the Li stellar enrichment, with mass loss values more than one thousand times that of the normal mass loss of a K-giant star, only the inner parts of the debris disks are eroded. 
This dragging process removes particles less than some centimeters in size, leaving larger bodies unaltered. All these small particles, at least those in the micron-sized scale, remain in the expanding shell during these very rapid stages observed in these emerging new Li stars. The existence of these micron-sized particles is, in fact, inferred by the observation of the continuum emission spectra in all the RGB stars. 
Also the presence of micron-sized particles is observed in a direct case of shell-disk interaction. This is the case 
of one of the super Li rich RGB star, where micron sized crystalline Enstatite particles, typical inner debris disk particles, are injected into the shell. 
We also expect that this wind-disk interaction can, in principle, form non-spherical geometries that could be detected with ALMA. Furthermore, optical polarimetric observations showed that at least two stars (one is the star showing the Enstatite signatures) of these emerging very Li rich giants exhibit polarisation signals. 

We can conclude that, if this episodic Li enrichment process is indeed an universal phenomenon as evidenced by Galactic and extragalactic observations, we are in face of a new source of organics in the interstellar medium.

\acknowledgements The authors thank the anonymous referee for comments that helped improve the paper. 
This work is based on Spitzer observations processed through CASSIS, for which the authors are grateful. 
N.A.D. thanks Saint Petersburg State University, Russia, for research grant 6.38.18.2014 and 
FAPERJ, Rio de Janeiro - Brazil, for Visiting Researcher grant E-26/200.128/2015.
I.O. is supported by a level A Young Talent grant under the Science Without Borders program of CNPq - Brazil.

\clearpage

\clearpage

\clearpage

\clearpage

\clearpage

\clearpage

\begin{deluxetable}{lcccccl}
\tabletypesize{\tiny}
\rotate
\tablecaption{Stellar properties and spectral emission features}
\tablewidth{0pt}
\tablehead{
\colhead{Star} & \colhead{Number} & \colhead{Evolution} & \colhead{$T_{\rm eff}$. K} & \colhead{$\log\epsilon$(Li)} &
\colhead{$\log\,L/L_{\odot}$} & \colhead{Emission features ($\micron$)}  \\
\colhead{} &\colhead{in Fig.\,1}  &\colhead{Phase}& \colhead{}  &\colhead{}   &\colhead{}    &\colhead{}         \\
}
\startdata
HD\,233517    & 15       & RGB    & 4390   & 4.3            & 2.03$^a$    &  6.26$^b$  A, 8.2  ALP?,  11.3$^b$ A,  12.7 ALP?    \\    
PDS\,365      & 30       & RGB    & 4540   & 3.3            & 1.86$^c$    & 6.26 A,  8 - 9 ALP,  20.0 F/E?,  34.0 F/E?
       \\  
IRAS\,17596-3952& 62     & RGB    & 4600   & 2.30           & 1.70$^{a,d}$ & 8.3$^e$ Si?,  20.0 F/E ?, 34.0 F/E ?   \\
PDS\,100      & 74       & RGB    & 4500   & 2.40           & 1.65$^{a,f}$ & 6.26 A, 11.4 A,  19.0 F/E?,  23.5 F/E?,  28.0 F/E?,  33.0 UIE  \\ 
PDS\,68       & 34       & RGB    & 4300   & 3.9 - 4.2      & 1.60$^{a,d}$ & 10.0$^g$ CE, 12.7 ALP,  19.5$^g$  CE,  28.3$^g$  CE  \\
IRAS\,17582-2619 & 60    & EAGB?  & -      & $\sim$1.40     &      -       & 8.7 ALP \\  
IRAS\,12327-6523$^{(b)}$ & 27 & EAGB & 4200 & 1.4 -- 1.6    & 2.91$^{d}$   & 10.0  AM, 18.5 AM  \\  
\hline
\enddata
\tablecomments{Abbreviations: A  - aromatic, ALP - aliphatic plateau F - forsterite,  E- enstatite, Si - silica, 
               CE - crystal enstatite, AM - amorphous inorganic compound, UIE - unidentified infrared emission}
\tablerefs{(a) \citet{stra15}; (b) \citet{jur06}; (c) \citet{dra02}; (d) \citet{red05}; (e) \citet{juh10}; (f) \citet{red02}; (g) \citet{olo09}.}
\end{deluxetable}

\end{document}